# LMU-Based Sequential Learning and Posterior Ensemble Fusion for Cross-Domain Infant Cry Classification

Niloofar Jazaeri
*University of Ottawa*
Ottawa, Canada
njaza024@uottawa.ca

Hilmi R. Dajani
*University of Ottawa*
Ottawa, Canada
hdajani@uottawa.ca

Marco Janeczek
*Crynostics Inc.*
Canada
marco@crynostics.com

Martin Bouchard
*University of Ottawa*
Ottawa, Canada
bouchm@uottawa.ca

***Abstract*—Decoding infant cry causes remains challenging for healthcare monitoring due to short nonstationary signals, limited annotations, and strong domain shifts across infants and datasets. We propose a compact acoustic framework that fuses mel-frequency cepstral coefficients (MFCCs), short-time Fourier transform (STFT) features, and fundamental-frequency ($F_0$) contours within a multi-branch convolutional neural network (CNN) encoder, and models temporal dynamics using an enhanced Legendre Memory Unit (LMU). Compared to LSTMs, the LMU backbone provides stable sequence modeling with substantially fewer recurrent parameters, supporting efficient deployment. To improve cross-dataset generalization, we introduce calibrated posterior ensemble fusion with entropy-gated weighting to preserve domain-specific expertise while mitigating dataset bias. Experiments on Baby2020 and Baby_Crying demonstrate improved macro-F1 under cross-domain evaluation, along with leakage-aware splits and real-time feasibility for on-device monitoring.**

***Index Terms*—Infant cry, Legendre memory unit, sequential learning, ensemble learning, posterior fusion, domain adaptation.**

## I. Introduction

Infant cry is one of the earliest and most informative communication signals between newborns and caregivers. Accurately decoding cry causes (e.g., hunger, discomfort, pain) can improve parental responsiveness and provide non-invasive clinical cues for early detection of pathological conditions [1]–[3]. However, even experienced caregivers struggle to reliably distinguish acoustic nuances, and expert annotation is costly and inconsistent, motivating machine listening systems that can robustly interpret infant cries across diverse settings.

Despite progress, cry analysis remains challenging. Cries are short, nonstationary, and highly variable across infants and sessions. The datasets are small and imbalanced, and they are prone to overestimation due to *leakage*—where portions of segments or augmented cry samples from the same cry appear in both the training and test sets. Environmental variability such as background noise, overlapping speech, and sounds from television further undermines generalization. Prior work on cry detection in naturalistic conditions [4] and infant monitoring systems [5] highlighted how acoustic interference severely reduces system reliability, underscoring the need for models resilient to domain shift. Cry-cause classification is particularly sensitive, as it requires capturing fine-grained spectral and prosodic dynamics beyond simple cry/non-cry detection.

A broad spectrum of methodologies has been explored in prior work. Early efforts used classical machine learning such as SVMs, Random Forests, and ensemble trees on handcrafted features [6], [7], sometimes combined with transfer learning for robustness. Multi-stream classifiers have improved performance by fusing complementary acoustic views [8], [9], but most ensembles operate at the feature or decision level without addressing domain mismatch. Deep learning has since become dominant: CNNs, RNNs, and CNN–LSTM hybrids model spectro-temporal dependencies [10], [11], while ResNet and EfficientNet backbones have been benchmarked for infant cry detection [12]. More advanced approaches exploit graph neural networks [13], attention-based layers [14], ontology-driven representation learning [15], or Whisper encoder features [16]. Recent surveys highlight contrastive and multi-stream pipelines as emerging trends in cry representation learning [11]. However, while ensemble methods have been used to combine modalities or feature sets, they have not been systematically explored for *domain adaptation within the same acoustic modality*, where differences in labeling practices and acoustic environments create cross-dataset inconsistencies.

Parallel advances in self-supervised and contextualized audio representation learning further broaden the landscape. Conformer-based contrastive learning [17], contextualized acoustic embeddings [18], and compositional or contrastive cry representations [19] show promise for leveraging unlabeled audio. Domain-specific embeddings, such as neonatal health monitoring pipelines, point toward clinically viable applications, while multi-microphone separation [20] strengthens robustness under noisy conditions.

While LSTMs and GRUs remain a dominant recurrent backbone, their reliance on multiple gates makes them parameter-heavy and prone to training instabilities over long

This work was supported by a Discovery Grant from the Natural Sciences and Engineering Research Council of Canada (NSERC) and by a MITACS Accelerate internship.

sequences [21]. The recently proposed *Legendre Memory Unit* (LMU) [22], by contrast, formulates recurrent memory as a state-space projection onto orthogonal Legendre polynomials. LMUs propagate gradients stably, provide explicit control over memory span, and require up to an order of magnitude fewer parameters than LSTMs and GRUs, making them particularly suitable for lightweight, on-device cry analysis.

**In this work, we make four key contributions.** (1) We introduce a compact, time-preserving encoder fused with an LMU sequence model, using established acoustic and sequential modeling components to achieve comparable or superior performance to LSTMs with significantly fewer recurrent parameters. (2) We establish a *leakage-aware evaluation protocol* to ensure unbiased training and testing across infant datasets. (3) We propose a *calibrated posterior fusion ensemble* with entropy-weighted averaging as the key novel component of this work; this posterior-level domain adaptation strategy explicitly addresses cross-dataset inconsistencies by preserving dataset-specific expertise, handling partially overlapping label spaces, and retaining minority-class information. (4) We validate the resulting framework for *real-time mobile deployment*, showing that lightweight models (~5 MB) run within ~3 s per 10 s cry clip, demonstrating feasibility for on-device pediatric monitoring. In summary, the main original contribution is the calibrated posterior fusion ensemble for cross-domain infant cry classification, while the feature extraction pipeline, CNN encoder, and LMU/LSTM sequential models are established components used to build an efficient end-to-end framework.

## II. Datasets and Ethics

We evaluate our framework on two publicly available infant cry corpora. Following critiques of overestimation due to sample leakage [4], [5], all splits are leakage-safe, ensuring no baby or session overlap across training, validation, and test sets. A summary of the key properties of both datasets is provided in Table I.

**Baby2020 [13].** We use the 0–3 month subset, containing 2,190 recordings (1–7 s, 16 kHz) labeled as *hungry* (590), *sleepy* (1,000), or *awake* (600). Recordings follow a structured naming convention that encodes infant age/sex and segment indices, enabling leakage-aware grouping. Parents provided continuous recordings, later segmented into labeled events by Ji *et al.* [13]. This dataset has become a benchmark for structured cry modeling in CNN+GNN and attention-based pipelines [13], [14].

**Baby_Crying [23].** This dataset contains 371 recordings with more diverse acoustic conditions: durations range from 7–30 s, sampling rates vary, and background noise is moderate. Classes include *awake* (128), *uncomfortable* (128), *diaper* (134), *hungry* (160), and *sleepy* (115). Only session-level IDs are available [11].

**Ethics statement.** This study used only publicly available, de-identified infant cry datasets and involved no new human-subject or animal experiments. The datasets were collected and released by their original investigators under their reported ethics/consent protocols; no additional institutional approval was required.

TABLE I
Summary of datasets used in this study (leakage-free splits).

| **Property** | **Baby2020** | **Baby_Crying** |
|---|---|---|
| Recordings | 2,190 | 371 |
| Duration (s) | 1–7 | 7–30 (avg. ~20) |
| Noise level | Low | Moderate |
| Sample rate | 16 kHz | Resampled to 16 kHz |
| ID granularity | Baby+Session | Session |
| Classes | 3 | 5 |

## III. Methodology

### A. Feature Extraction

We extract four complementary acoustic representations—MFCCs, STFT log-power spectra, $F_0$ contours with confidence, and waveform energy—to capture spectral, cepstral, prosodic, and amplitude cues. Their joint utility is confirmed in ablation studies (§V).

**MFCC.** We compute $D$=13 mel-frequency cepstral coefficients (MFCCs) per frame using a 30 ms Hamming window and 10 ms hop size. Let $X(k, n)$ denote the short-time Fourier transform (STFT) of the signal, where $k \in \{0, \dots, K-1\}$ indexes discrete DFT frequency bins and $n \in \{1, \dots, N\}$ indexes analysis frames. The power spectrum $|X(k, n)|^2$ is passed through $M$ triangular mel filters banks $\{H_m(k)\}_{m=1}^{M}$, each emphasizing energy within a specific mel-spaced frequency band. The energy in the $m$-th mel band at frame $n$ is computed as

$$E_m(n) = \sum_{k=0}^{K-1} |X(k, n)|^2 \, H_m(k), \tag{1}$$

where $E_m(n)$ represents the total spectral energy captured by the $m$-th mel filter. Applying a discrete cosine transform (DCT) to the log energies yields the cepstral coefficients:

$$c_d(n) = \sum_{m=0}^{M-1} \log E_m(n) \cos\left[\frac{\pi d}{M}(m + 0.5)\right], \tag{2}$$

with $d \in \{0, \dots, D-1\}$ denoting the cepstral index. The MFCCs thus compactly encode the smooth spectral envelope of each frame, providing robustness to noise and minor channel variations by discarding fine-scale harmonic structure.

**STFT.** Log-power STFTs are computed with a 512-point FFT, 30 ms window, and 50% overlap:

$$S(k, n) = \log\left(|X(k, n)|^2 + \epsilon\right), \tag{3}$$

where $\epsilon$ is a small positive constant ($\epsilon = 10^{-10}$ ) added for numerical stability to avoid the logarithm of zero.

**$F_0$ and confidence.** We extract $F_0$ contours and confidence $c(n) \in [0, 1]$ using the Convolutional REpresentation for Pitch Estimation (CREPE) model [24]:

$$\hat{f}_0(n), \, c(n) = \text{CREPE}(x[n]). \tag{4}$$

Frames with low $c(n)$ are down-weighted. Prosodic cues from F0 trajectories help distinguish urgency in cries.

**Waveform energy.** The raw waveform $x[n]$, loaded using `librosa.load`, is used at its native floating-point amplitude scale (typically bounded in $[-1, 1]$ for PCM audio) without explicit peak or RMS normalization. The waveform is resampled in time to a fixed length of $T$ using linear interpolation and provides amplitude and rhythmic envelope cues that complement the spectral features.

**Feature fusion.** Because different acoustic features (MFCC, STFT, $F_0$ with confidence, and waveform-based descriptors) are computed on distinct temporal grids, all feature matrices are resampled to a common frame length of $T$. Selecting the median frame length balances information preservation and memory efficiency—avoiding excessive zero-padding (as with the maximum length) and minimizing distortion from trimming shorter clips.

Let $F_m \in \mathbb{R}^{D_m \times T_m}$ denote the feature matrix for modality $m$, where $D_m$ is the number of feature coefficients and $T_m$ is the original number of temporal frames. Each matrix $F_m$ is linearly interpolated along the temporal axis to the unified length $T$. For each target frame $t \in \{1, \ldots, T\}$, let

$$s_t = 1 + (t-1)\frac{T_m - 1}{T - 1}, \quad i_t = \lfloor s_t \rfloor, \quad \lambda_t = s_t - i_t. \quad (5)$$

Then, for each feature coefficient $d \in \{1, \ldots, D_m\}$, the resampled value is

$$\tilde{F}_m(d, t) = (1 - \lambda_t)F_m(d, i_t) + \lambda_t F_m(d, i_t + 1), \quad (6)$$

with boundary clipping when $i_t = T_m$. This explicitly aligns all modalities to a shared temporal grid while preserving feature-wise temporal dynamics.

**Time alignment (median length).** The common length was set to $T$=233, the median frame count of the training set, to reduce zero-padding and avoid aggressive truncation. The aligned features are concatenated along the feature axis, producing a $(273 \times 233)$ tensor consisting of 257 STFT bins, 13 MFCCs, $F_0$, $F_0$ confidence, and waveform energy.

### *B. Encoder & Feature Fusion*

Each cry sample is represented as a 233×273 matrix. A CNN stem with three Convolutional - batch normalization - pooling blocks (128, 64, 32 filters) extracts spectro-temporal patterns, following prior cry spectrogram pipelines [11], [13]. Kernel sizes capture both fine harmonics and broader shifts, with dropout and L2 regularization reducing overfitting. The output is flattened per frame to preserve temporal order for sequence modeling.

### *C. Sequential Memory Models: LSTM vs. LMU*

LMU was selected because infant cries are short but temporally structured signals whose spectral and prosodic trajectories evolve over several hundred frames. Its fixed orthogonal memory dynamics provide stable temporal modeling with fewer recurrent parameters than gated RNNs, which is important for deployment-oriented infant monitoring. LSTM was used as the primary recurrent baseline because it remains a widely used sequence model in audio classification, while GRU, CNN-only, and Transformer baselines are included in Section IV to compare against both simpler and more recent alternatives.

**Legendre Memory Unit (LMU).** The LMU [22] implements recurrent memory as a continuous-time state-space system that projects recent input samples onto a fixed set of orthogonal Legendre polynomial basis functions over a temporal window $\theta$. At each discrete timestep $t$, the network receives an input vector $x_t \in \mathbb{R}^p$ and maintains a linear memory state $m_t \in \mathbb{R}^d$, where $d$ denotes the *memory order* (i.e., the number of Legendre basis functions representing the recent history). The hidden activation (readout) is $h_t \in \mathbb{R}^r$.

The discretized LMU dynamics are defined as:

$$m_{t+1} = \mathbf{A}\, m_t + \mathbf{B}\, u_t, \quad (7)$$

$$h_t = \mathbf{C}\, m_t + \mathbf{D}\, u_t, \quad (8)$$

where $(\mathbf{A}, \mathbf{B}, \mathbf{C}, \mathbf{D})$ are fixed matrices derived from the continuous Legendre basis and the window length $\theta$. These matrices encode how recent input samples are projected into the orthogonal memory subspace, providing a stable and well-conditioned temporal representation.

The nonlinear input projection combines the current external input, the previous hidden activation, and the previous memory state:

$$u_t = W_x x_t + W_h h_{t-1} + W_m m_{t-1}, \quad (9)$$

where $W_x \in \mathbb{R}^{q \times p}$ maps the input features, $W_h \in \mathbb{R}^{q \times r}$ maps the recurrent hidden activations, and $W_m \in \mathbb{R}^{q \times d}$ maps the previous memory state into the input subspace. The resulting $u_t \in \mathbb{R}^q$ may optionally be passed through a nonlinearity such as $\tanh$ function before driving the memory update. This hybrid formulation—linear memory with nonlinear input mixing—preserves long-range temporal dependencies with *order-of-magnitude fewer parameters* than gated RNNs (e.g., LSTMs and GRUs) while maintaining stable gradients over extended time horizons.

**Long Short-Term Memory (LSTM).** LSTMs [21] regulate information flow using three gating mechanisms—*input*, *forget*, and *output* gates—that control how new information is integrated, retained, or exposed at each timestep $t$. Given an input vector $x_t \in \mathbb{R}^p$ and the previous hidden activation $h_{t-1} \in \mathbb{R}^r$, the gate activations and cell updates are computed as:

$$i_t = \sigma(W_i x_t + U_i h_{t-1} + b_i), \quad (10)$$

$$f_t = \sigma(W_f x_t + U_f h_{t-1} + b_f), \quad (11)$$

$$o_t = \sigma(W_o x_t + U_o h_{t-1} + b_o), \quad (12)$$

$$g_t = \tanh(W_c x_t + U_c h_{t-1} + b_c), \quad (13)$$

$$c_t = f_t \odot c_{t-1} + i_t \odot g_t, \quad (14)$$

$$h_t = o_t \odot \tanh(c_t). \quad (15)$$

Here, $\sigma(\cdot)$ denotes the logistic sigmoid function $\sigma(z) = 1/(1 + e^{-z})$ that squashes values to $[0, 1]$ and determines gate activations, while $\tanh(\cdot)$ introduces bounded nonlinearity in the cell update. The operator $\odot$ denotes element-wise (Hadamard) multiplication. Each $W_i, W_f, W_o, W_c \in \mathbb{R}^{r\times p}$ and $U_i, U_f, U_o, U_c \in \mathbb{R}^{r\times r}$ are weight matrices mapping input and recurrent signals, respectively, and $b_i, b_f, b_o, b_c \in \mathbb{R}^r$ are bias vectors. The internal state $c_t \in \mathbb{R}^r$ acts as the long-term memory, while $h_t \in \mathbb{R}^r$ represents the short-term (hidden) activation used for output and recurrence.

Each update involves multiple matrix multiplications, nonlinear activations, and gating operations—making LSTMs highly expressive but relatively heavy in parameters and computational cost.

**Comparison.** Unlike LSTMs, which learn all recurrent weights, LMUs employ fixed orthogonal memory dynamics with a learned readout layer, achieving approximately 95% fewer recurrent parameters [22]. This design yields faster training, lower inference latency, and improved numerical stability, making LMUs particularly suitable for mobile and real-time deployment, whereas LSTMs remain competitive for strongly nonlinear or event-driven temporal dynamics.

### *D. Domain Adaptation via Calibrated Posterior Fusion*

We train two domain-specific classifiers: one on Baby2020 with labels *hungry*, *awake*, *sleepy* and one on Baby_Crying with labels *hug*, *uncomfortable*, *sleepy*. At inference, their outputs are projected into a shared union label space $\mathcal{L}$ =*hungry*, *awake*, *hug*, *uncomfortable*, *sleepy*, as shown in the workflow of Fig. 1.

**Algorithmic steps.** Algorithm 1 summarizes the proposed calibrated posterior fusion procedure. Each domain-specific classifier is first temperature-calibrated on its validation set by learning a positive scalar $T_m$ that minimizes validation negative log-likelihood, correcting overconfident logits before softmax.

During inference, calibrated logits are projected into the shared union label space. Disjoint classes are inserted directly, while the shared class *sleepy* is fused using entropy-weighted log-sum-exp, so lower-entropy models contribute more strongly. An optional product-of-experts variant was also evaluated for sharper consensus when both models agree.

**Illustrative examples of calibrated posterior fusion.** We next highlight how calibration and entropy-gated fusion resolve conflicts across datasets.

**Interpretation of calibrated posterior fusion behavior.** Table II presents representative, sample-level case studies illustrating how calibrated posterior fusion combines domain-specific predictions from the Baby2020 and Baby_Crying models. Unlike the aggregate metrics reported in Section IV, these examples serve a qualitative and explanatory role, clarifying *how* temperature calibration and entropy-gated weighting influence individual decisions.

**Why ensemble domain adaptation matters for baby cry.** Compared to naive dataset merging, posterior fusion preserves domain-specific strengths and reduces bias. In cry classification, merging datasets can worsen class imbalance and obscure annotation differences, while majority voting fails

**Algorithm 1** Leakage-safe training and calibrated posterior fusion across Baby2020 and Baby_Crying datasets.

**Require:** Baby2020 dataset $\mathcal{D}_B$ with labels $\{\text{hungry}, \text{awake}, \text{sleepy}\}$.
**Require:** Baby_Crying dataset $\mathcal{D}_C$ with labels $\{\text{hug}, \text{uncomfortable}, \text{sleepy}\}$.
**Require:** Union label space
$\mathcal{L} = \{\text{hungry}, \text{awake}, \text{hug}, \text{uncomfortable}, \text{sleepy}\}$.

**Stage 1: Leakage-safe training**
1: **for** $\mathcal{D}_m \in \{\mathcal{D}_B, \mathcal{D}_C\}$ **do**
2: Split $\mathcal{D}_m$ into train, validation, and test sets with no baby/session overlap.
3: Train CNN–LMU classifier $f_m$ with early stopping on validation macro-F1.
4: **end for**

**Stage 2: Post-hoc temperature calibration**
5: **for** each trained classifier $f_m$ **do**
6: Collect validation logits $z_m^{(n)}$ and labels $y^{(n)}$.
7: Estimate the calibration temperature:

$$T_m = \arg\min_{T>0} \left[ -\sum_n \log \frac{\exp(z_m^{(n)}[y^{(n)}]/T)}{\sum_{j\in\mathcal{Y}_m} \exp(z_m^{(n)}[j]/T)} \right]. \tag{16}$$

8: **end for**

**Stage 3: Calibrated posterior fusion**
9: **for** each test cry recording $x$ **do**
10: Compute calibrated posterior probabilities:

$$P_m(y|x) = \frac{\exp(z_m(y|x)/T_m)}{\sum_{j\in\mathcal{Y}_m} \exp(z_m(j|x)/T_m)}. \tag{17}$$

11: Compute entropy-gated model weights:

$$w_m = \frac{\exp[-\tau H(P_m)]}{\sum_{\ell\in\{B,C\}} \exp[-\tau H(P_\ell)]}. \tag{18}$$

12: Initialize union logits $z_{\mathcal{L}} \in \mathbb{R}^{|\mathcal{L}|}$ with $-\infty$.
13: Insert logits for non-overlapping classes directly into $z_{\mathcal{L}}$.
14: Fuse the shared class *sleepy* by weighted log-sum-exp:

$$z_{\mathcal{L}}[\text{sleepy}] = \log\left(w_B \exp(z_B[\text{sleepy}]/T_B) + w_C \exp(z_C[\text{sleepy}]/T_C)\right). \tag{19}$$

15: Compute the final fused prediction:

$$p^\star(y|x) = \frac{\exp(z_{\mathcal{L}}[y])}{\sum_{j\in\mathcal{L}} \exp(z_{\mathcal{L}}[j])}, \qquad \hat{y} = \arg\max_{y\in\mathcal{L}} p^\star(y|x). \tag{20}$$

16: **end for**

when label spaces differ (e.g., *hungry* in Baby_Crying vs. *hug* in Baby2020). Treating each dataset as an expert and combining them only at inference through calibrated posterior fusion avoids annotation conflicts, improves minority-class representation, and broadens domain coverage without retraining. Table III quantitatively compares these domain adaptation strategies under cross-dataset evaluation.

Each case contrasts uncalibrated and calibrated fusion outcomes. A case is considered *successful* when both domains agree on the correct label, or when calibration and entropy gating resolve cross-domain conflicts by reducing overconfidence or prioritizing the more reliable expert. Conversely, a case is *unsuccessful* when a domain-specific model dominates the ensemble due to unjustified low entropy, causing misclassification.

The first example (*Hungry – calibration effect*) shows that temperature scaling ($T_C$=1.6, $T_B$=0.8) softens overconfident Baby_Crying posteriors, enabling balanced fusion while preserving the correct *Hungry* prediction. The second case demonstrates entropy gating, where the lower-entropy Baby_Crying prediction is weighted more strongly when con-fident and correct. In the *Hug* case, Baby2020 provides a sharp low-entropy posterior that overrides the diffuse Baby_Crying output, while the *Sleepy* case shows straightforward consensus when both domains agree.

TABLE II
REPRESENTATIVE CASE STUDIES OF CALIBRATED POSTERIOR FUSION ACROSS BABY2020 AND BABY_CRYING MODELS. MOST CASES ILLUSTRATE SUCCESSFUL CONFLICT RESOLUTION VIA CALIBRATION AND ENTROPY GATING; THE FINAL ROW HIGHLIGHTS A FAILURE CASE WHERE A HIGHLY CONFIDENT BUT INCORRECT DOMAIN PREDICTION DOMINATES.

| Case | Posterior behavior and fusion outcome |
|---|---|
| **Cry: Hungry (calibration effect)** | Baby_Crying: hungry=0.92, sleepy=0.08. Baby2020: hug=0.60, sleepy=0.40. Without calibration: predicts **hungry** (overconfident). With calibration ($T_C$=1.6, $T_B$=0.8): hungry=0.70 vs. hug=0.65 $\Rightarrow$ **hungry**. |
| **Cry: Hungry (entropy gating)** | Baby_Crying: hungry=0.88, sleepy=0.08, awake=0.04. Baby2020: sleepy=0.55, discomfort=0.45. Lower entropy in Baby_Crying $\Rightarrow$ fusion favors **hungry**. |
| **Cry: Hug** | Baby_Crying: diffuse across hungry/awake/sleepy. Baby2020: hug=0.92 (low entropy). Fusion correctly favors Baby2020 $\Rightarrow$ **hug**. |
| **Cry: Sleepy** | Baby_Crying: sleepy$\sim$0.8; Baby2020: sleepy$\sim$0.8. Agreement across domains $\Rightarrow$ **sleepy**. |
| **Cry: Sleepy (failure case predicted Hungry)** | Baby_Crying: hungry=0.95 (very low entropy, incorrect label). Baby2020: sleepy=0.60, hug=0.40 (higher entropy, correct). Entropy gating strongly weights Baby_Crying, causing fusion to incorrectly predict as **hungry**. |

The final row of Table II highlights a limitation: when a domain-specific model is confidently wrong, its low-entropy posterior can dominate the ensemble and override a more uncertain but correct prediction. This motivates uncertainty-aware rejection and dynamic domain reliability estimation.

Overall, calibrated posterior fusion improves reliability by tempering overconfident probabilities, weighting domain contributions by predictive entropy, and producing more consistent cross-domain decisions. These qualitative examples complement the quantitative evaluation in Section IV by revealing the internal decision behavior of the proposed fusion strategy.

**Why this works.** This calibrated fusion respects disjoint labels, combines overlap fairly, and prevents overconfident models from dominating. It allows domain expertise to be preserved without retraining, while providing flexible control through temperature scaling and entropy-based gating.

## IV. EXPERIMENTS

**Cry detection front-end.** We adopt the cry/non-cry detector of Yao *et al.* [4], retrained and re-saved in TensorFlow 2.x for compatibility with modern Python environments (>3.8). The detector operates on 1 s windows with 80% overlap and rejects non-cry events such as adult speech, background television, and silence (Fig. 2). Its footprint is $\sim$300 MB, with average latency of $\sim$3 s per 10 s audio clip on an AWS CPU instance.

This module is *not required* for curated datasets such as Baby2020 or Baby_Crying, which already contain pre-segmented cry samples. However, in real-world deployment, continuous audio often includes noisy mixtures of infant and non-infant sounds. The detection stage filters irrelevant segments and isolates cry episodes, reducing the gap between clean training data and field conditions.

**Parameter selection.** Feature-extraction parameters were selected to balance temporal resolution, spectral detail, and computational cost. The 30 ms window and 10–15 ms hop sizes follow standard short-time audio analysis practice, while the 512-point FFT provides sufficient spectral resolution at 16 kHz with modest memory cost. The unified temporal length $T$=233 was set to the median training-set frame count to reduce zero-padding and avoid aggressive truncation. Model hyperparameters (CNN filters, dropout, learning rate, batch size, and early stopping) were selected using validation macro-F1 only; the test set was held out for final evaluation.

**Train–validation–test split.** All experiments used leakage-aware group-level splits rather than segment-level splits. We did not use cross-validation; instead, results were averaged over five independent random-seed splits. For each split, 20% of the data was held out for testing and 80% for training, with 10% of the training portion reserved for validation. When baby or session identifiers were available, splitting was performed at that level to reduce leakage. The validation set was used for early stopping, hyperparameter selection, and temperature calibration; the test set was used only for final evaluation.

**Training setup.** The *Baby2020* classifier was trained for up to 1000 epochs with batch size 8 and learning rate $10^{-3}$ on 7×10 GB GPUs; the *Baby_Crying* classifier used batch size 32. Early stopping used validation macro-F1, and mixed precision reduced memory usage. The final classifier is compact ($\sim$5 MB) and suitable for mobile deployment.

**Feature ablation.** We evaluated $F_0$ contours from CREPE with confidence [24], STFT, and MFCC features extracted per cry segment, resampled to a common time axis and z-scored before training. As shown in Table IV, MFCC+STFT achieves the highest Macro-F1 on Baby_Crying (Diaper, Sleepy, Uncomfortable), while $F_0$ alone is substantially less discriminative, suggesting that timbral and spectral cues dominate these categories.

Pitch contribution is more dataset-dependent. On Baby2020, Table V shows that MFCC+STFT+$F_0$ gives the best F1, indicating that prosodic cues help under more structured recording conditions. Waveform energy was excluded from the Baby_Crying ablation to isolate pitch effects and reveal cases where $F_0$ estimation is unreliable. These results motivate retaining the full MFCC+STFT+$F_0$ feature set for a unified cross-dataset and deployment-ready pipeline.

**Sequential backbone comparison.** Table VI compares sequential learning models on fused features. Our CNN+LMU achieves the best trade-off, outperforming LSTM/GRU in both accuracy and efficiency, with $\sim$95% fewer recurrent parameters than LSTM [22]. Compared to graph-based models [13], [14], LMU matches or exceeds performance while training faster, supporting real-time usability.

**Interpretation.** SOTA (*state-of-the-art*) baselines from

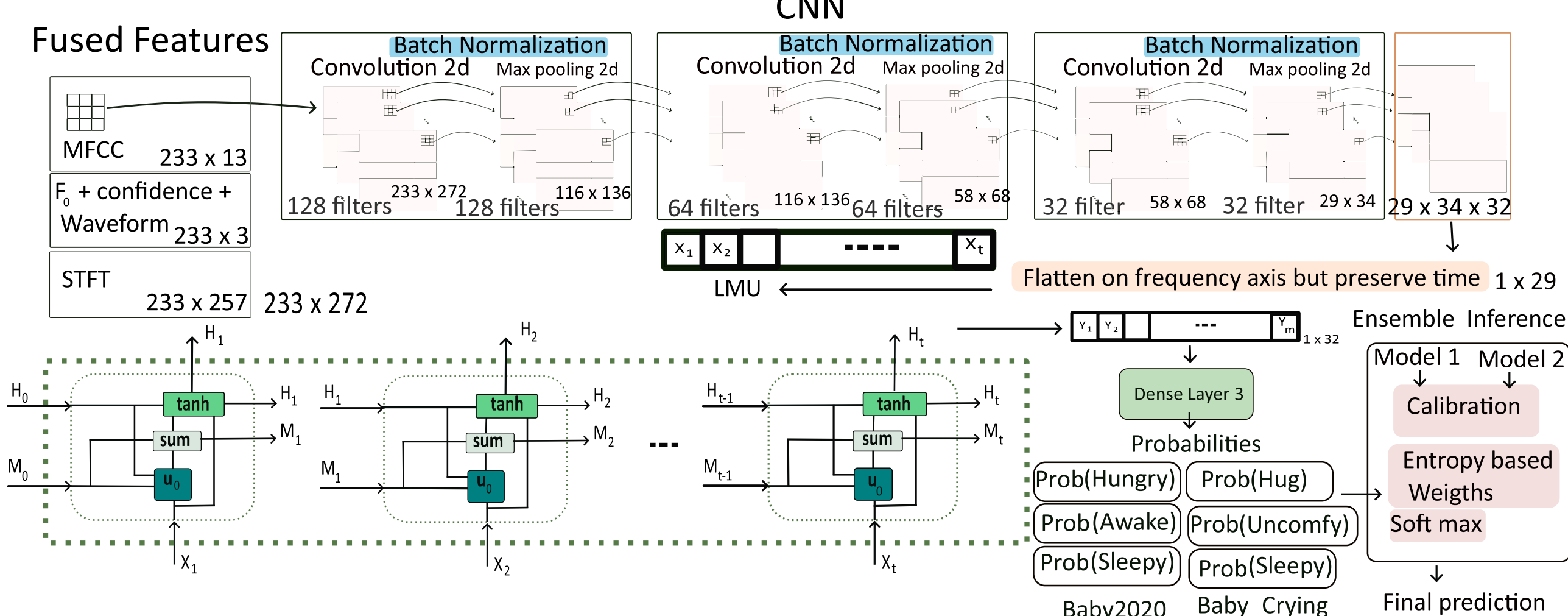

Fig. 1. Block diagram of the proposed ensemble learning approach for infant cry cause classification. Each branch integrates feature selection and sequential learning layers, and their outputs are combined using calibrated ensemble learning to enhance robustness.

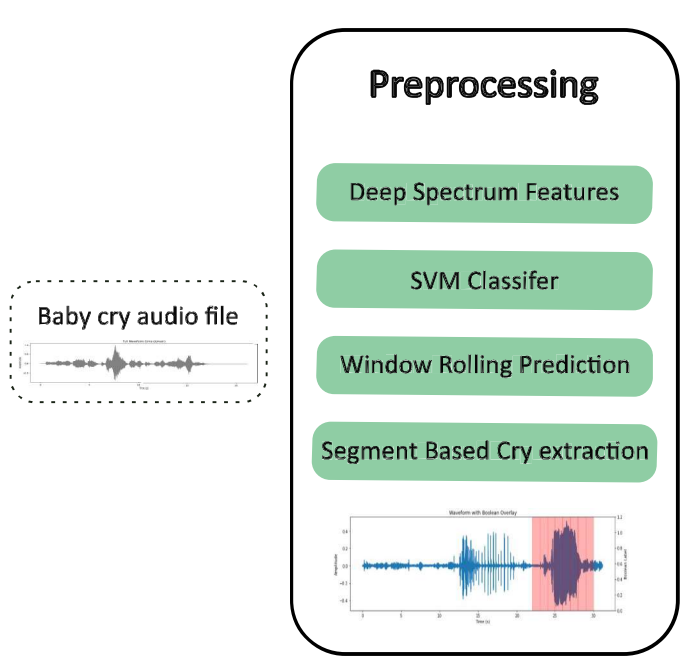

Fig. 2. Infant cry detection workflow. Non-cry segments (e.g., adult speech, middle red ribbon) are rejected. Remaining cry segments are classified into mood categories via an RBF SVM.

TABLE III

COMPARISON OF DOMAIN ADAPTATION STRATEGIES ACROSS BABY2020 AND BABY_CRYING DATASETS. RESULTS ARE MACRO-F1 (MEAN ± STD OVER 5 RANDOM SEEDS). PROPOSED CALIBRATED FUSION YIELDS BEST AVERAGE CROSS-DOMAIN GENERALIZATION.

| Method | Test: Baby2020 | Test: Baby_Crying | Avg. Test |
|---|---|---|---|
| Single (Baby_Crying→Baby2020) | 0.27±0.02 | 0.78±0.03 | 0.53±0.03 |
| Single (Baby2020→Baby_Crying) | 0.76±0.03 | 0.26±0.03 | 0.51±0.03 |
| Majority Vote (Agree) | N/A | N/A | N/A |
| SoftAvg (Uncalibrated) | 0.61±0.04 | 0.60±0.03 | 0.61±0.04 |
| **Calibrated Fusion (ours)** | 0.62±0.04 | 0.65±0.03 | **0.64±0.04** |
| Merged (Joint Training) | 0.58±0.05 | 0.55±0.04 | 0.57±0.05 |

prior literature [14], [13] represent the best published Baby2020 models using graph-based or attention-enhanced CNNs. For a fair internal comparison, we implemented LSTM, GRU, Transformer, and LMU backbones on the same CNN encoder with comparable capacity. The Transformer baseline

TABLE IV

FEATURE ABLATION ON THE BABY_CRYING DATASET (DIAPER, SLEEPY, UNCOMFORTABLE). MODELS ARE TRAINED WITH AN 80% / 20% TRAIN–TEST SPLIT CNN-LSTM MODEL. WE REPORT MACRO-F1 (%). MFCC+STFT YIELDS THE STRONGEST PERFORMANCE.

| Features | Macro-F1 (%) |
|---|---|
| MFCC + F0 + STFT | 79.5 |
| F0 only | 46.0 |
| MFCC only | 80.0 |
| STFT only | 82.0 |
| F0 + MFCC | 78.0 |
| F0 + STFT | 80.0 |
| **MFCC + STFT** | **85.0** |

TABLE V

ABLATION STUDY OF ACOUSTIC FEATURES ON BABY2020 DATASETS. METRIC: MACRO-F1 USING CNN+LSTM.

| Features | F1-Score |
|---|---|
| F0 only | 0.54 |
| STFT only | 0.60 |
| MFCC only | 0.58 |
| F0 + STFT + MFCC | **0.76** |

addresses recent attention-based alternatives, while LSTM provides a standard recurrent reference for audio sequence modeling; only optimization parameters such as learning rate, dropout, and regularization were tuned.

This controlled setup isolates the effect of the sequential learner rather than model size. The LMU achieves the best Macro-F1 among internal baselines, suggesting that its linear memory formulation captures smooth temporal cry dynamics while maintaining stable gradients and low parameter count. Compared with SOTA graph-based methods, CNN+LMU provides competitive cross-domain generalization with fewer

TABLE VI
COMPARISON OF SEQUENTIAL LEARNERS AND SOTA (*state-of-the-art*) BASELINES ON FUSED ACOUSTIC FEATURES ($F_0$ + STFT + MFCC). METRIC: MACRO-F1 . ALL SEQUENTIAL MODELS WERE IMPLEMENTED WITH IDENTICAL LAYER DEPTH AND NUMBER OF HIDDEN UNITS; ONLY TRAINING HYPERPARAMETERS WERE TUNED FOR FAIRNESS.

| Model | Dataset | Size | F1 |
|---|---|---|---|
| CNN | Baby2020 | 2,190 | 0.70 |
| CNN+LSTM | Baby2020 | 2,190 | 0.74 |
| CNN+GRU | Baby2020 | 2,190 | 0.71 |
| CNN+Transformer | Baby2020 | 2,190 | 0.67 |
| **CNN+LMU (ours)** | Baby2020 | 2,190 | **0.76** |
| AlgNet model and the efficient graph structure [14] | Baby_Crying | 918 | 0.93 |
| CRNN model and the efficient graph structure [14] | Baby_Crying | 918 | 0.91 |
| **CNN-LMU (ours)** | Baby_Crying | 371 | **0.85** |

parameters and faster convergence.

**Application inference.** The full pipeline was deployed on an AWS CPU server with an iOS Swift front-end. Latency was $\sim$3 s for cry detection and $\sim$3 s for classification per 10 s cry recording, well within real-time caregiver use. Model sizes (300 MB detector, 5 MB classifier) are lightweight enough for deployment, and quantization/pruning can reduce footprint further with minimal F1 drop ($< 0.5$ points).

## V. DISCUSSION AND CONCLUSION

We introduced a leakage-aware, domain-adaptive infant cry classification framework that fuses MFCC, STFT, and CREPE-based $F_0$ features with confidence, and employs an LMU backbone for efficient sequence modeling. LMUs provided well-conditioned memory for prosodic contours while CNN encoders captured local spectro-temporal cues, outperforming LSTM baselines with far fewer parameters and yielding lightweight models ($\sim$5 MB) deployable on mobile devices. Calibrated posterior fusion across Baby2020 and Baby_Crying classifiers mitigated dataset-specific bias and improved generalization under mismatched acoustic conditions. Remaining challenges include subjective labeling, limited age coverage, and "confidently wrong" cases, where a low-entropy but incorrect domain-specific prediction may dominate the fusion output. Future work will address these cases through uncertainty-aware rejection, class-wise calibration, dynamic domain reliability weighting, and broader age- and noise-diverse datasets. Overall, LMU-based modeling with calibrated cross-domain fusion offers an efficient and deployment-ready pathway for infant cry analysis.